\documentclass[aps,prd,twocolumn,groupeaddress,showpacs,showkeys,preprintnumbers]{revtex4-1}

\usepackage{epsfig}
\usepackage[usenames,dvipsnames]{color}

\usepackage{amsmath}
\usepackage{hyperref}

\begin{document}

\preprint{IPPP/17/5}

\title{Charming new physics in rare $B$-decays and mixing?}

\author{Sebastian J\"ager}
\affiliation{
  University of Sussex, Department of Physics and Astronomy,
  Falmer, Brighton BN1 9QH, UK}
\email{S.Jaeger@sussex.ac.uk}
\author{Matthew Kirk}
\affiliation{
  IPPP, Department of Physics, Durham University, Durham DH1 3LE, UK}
\email{m.j.kirk@durham.ac.uk}
\author{Alexander Lenz}
\affiliation{
  IPPP, Department of Physics, Durham University, Durham DH1 3LE, UK}
\email{alexander.lenz@durham.ac.uk}
\author{Kirsten Leslie}
\affiliation{
  University of Sussex, Department of Physics and Astronomy,
  Falmer, Brighton BN1 9QH, UK}
\email{K.Leslie@sussex.ac.uk}
\date{January, 2018}

\begin{abstract}
We conduct a systematic study of the impact of new physics
in quark-level $b \to c \bar  c s$ transitions on $B$ physics,
in particular rare $B$ decays and $B$-meson lifetime observables.
We find viable scenarios where a sizable effect in rare
semileptonic $B$ decays can be generated, compatible with experimental
indications and with a possible dependence on the dilepton invariant mass,
while being consistent with constraints from radiative $B$ decay and
the measured $B_s$ width difference. We show how, 
if the effect is generated at the weak scale or beyond, strong
renormalization-group effects can enhance the impact on semileptonic decays
while leaving radiative $B$ decay largely unaffected. A good
complementarity of the different $B$-physics
observables implies that precise measurements of lifetime observables
at LHCb may be able to confirm, refine, or rule out this scenario.

\end{abstract}


\maketitle

\section{Introduction}

Rare $B$ decays are excellent probes of new physics at the
electroweak scale and beyond, due to their strong suppression in
the Standard Model (SM). Interestingly, experimental data on rare branching
ratios \cite{Aaij:2014pli,Aaij:2015esa,*Khachatryan:2015isa}
and angular distributions for $B \to K^{(*)} \mu^+ \mu^-$ decay
\cite{Aaij:2015esa,*Khachatryan:2015isa,Lees:2015ymt,*Wei:2009zv,*Aaltonen:2011ja,*Aaij:2015oid,*Abdesselam:2016llu,*Wehle:2016yoi,*ATLAS:2017dlm,*Sirunyan:2017dhj}
may hint at a beyond-SM (BSM) contact interaction of the form
$
    (\bar s_L \gamma^\mu b_L ) (\bar \mu \gamma_\mu \mu)  ,
$
which would destructively interfere with the corresponding SM
(effective) coupling $C_9$ \cite{Descotes-Genon:2013wba,*Beaujean:2013soa,*Altmannshofer:2014rta,*Descotes-Genon:2015uva,*Hurth:2016fbr,Geng:2017svp,Altmannshofer:2017fio,*Ciuchini:2017mik,*Capdevila:2017bsm}, although
the significance of the effect is somewhat uncertain because
of form-factor
uncertainties as well as uncertain long-distance virtual charm
contributions \cite{Jager:2012uw,*Jager:2014rwa,*Ciuchini:2015qxb,*Chobanova:2017ghn,*Capdevila:2017ert,*Bobeth:2017vxj}.
However, if the BSM interpretation is correct, it requires
reducing $C_9$ by ${\cal O}(20\%)$ in magnitude. Such an effect might
arise from new particles (see e.g.\
\cite{Buras:2013qja,*Gauld:2013qja,*Buras:2013dea,*Datta:2013kja,*Altmannshofer:2014cfa,*Hiller:2014yaa,*Gripaios:2014tna,*Crivellin:2015mga,*Varzielas:2015iva,*Crivellin:2015lwa,*Becirevic:2015asa,*Celis:2015ara,*Alonso:2015sja,*Belanger:2015nma,*Falkowski:2015zwa,*Gripaios:2015gra,*Bauer:2015knc,*Fajfer:2015ycq,*Boucenna:2016qad,*Arnan:2016cpy,*Becirevic:2016yqi,*Crivellin:2016ejn,*GarciaGarcia:2016nvr,*Cline:2017aed,*Baek:2017sew,*Cline:2017ihf,*Kawamura:2017ecz,*DiChiara:2017cjq,*Kamenik:2017tnu,*Crivellin:2017zlb,*Ko:2017lzd,*Ko:2017yrd,*DiLuzio:2017vat}),
which might in turn be part of a more comprehensive
new dynamics.
Noting that in the SM, about half of $C_9$ comes from (short-distance)
virtual-charm contributions, in this article we ask whether new
physics affecting the quark-level $b \to c \bar c s$ transitions could
cause the anomalies, affecting rare $B$ decays
through a loop. The bulk of these effects would also be captured through an
effective shift $\Delta C_9(q^2)$, with a possible dependence on the
dilepton mass $q^2$.
At the same time,  such a scenario offers the exciting prospect of
confirming the rare $B$-decay anomalies through
correlated effects in hadronic $B$ decays into charm, with
``mixing'' observables such as the $B_s$-meson width difference
standing out as precisely measured \cite{Aad:2016tdj,*Khachatryan:2015nza,*Aaij:2014zsa,*Aaij:2014owa} and under reasonable theoretical
control. This is in contrast with the $Z'$ and leptoquark models usually
considered, where correlated effects are typically restricted to
other rare processes and are highly model dependent.  
Specific scenarios of hadronic new physics in the $B$ widths have been
considered previously
\cite{Bobeth:2014rda,*Bobeth:2014rra,*Brod:2014bfa,*Bauer:2010dga},
while the possibility of virtual charm BSM physics in rare
semileptonic decay has been raised in \cite{Lyon:2014hpa} (see also
\cite{He:2009hz}).
As we will show, viable scenarios exist, which
can mimic a shift $\Delta C_9 = -{\cal O}(1)$ while being consistent
with all other observables. In particular, very strong
renormalization-group effects can generate large shifts 
in the (low-energy) effective $C_9$ coupling from small $b \to c \bar c s$
couplings at a high scale without conflicting with the
measured  $\bar B \to X_s \gamma$ decay rate \cite{Chen:2001fja,*Abe:2001hk,*Aubert:2007my,*Limosani:2009qg,*Lees:2012ym,*Lees:2012ufa,*Lees:2012wg,*Saito:2014das}.

\section{Charming new physics scenario}
We consider a scenario where new physics affects the
$b \to c \bar c s$ transitions.
    This could be the case in models containing new scalars
    or new gauge bosons, or strongly coupled new physics. Such models
will typically affect other observables, but in a model-dependent manner.
For this paper, we restrict ourselves to studying
the new effects induced by modified  $b \to c \bar c s$ couplings,
leaving construction and phenomenology of concrete models for future work.
We refer to this as the ``charming BSM'' (CBSM) scenario.
As long as the mass scale $M$ of new physics satisfies $M \gg m_B$,
the modifications to the $b \to c \bar c s$ transitions
can be accounted for through a local effective Hamiltonian,
\begin{equation}\label{eq:heffcc}
    \mathcal{H}_{\rm eff}^{c\bar c}=\frac{4G_{F}}{\sqrt{2}} V_{cs}^*
    V_{cb}\sum_{i=1}^{10}(C_{i}^c Q^{c}_{i}+C^{c\prime}_{i} Q_i^{c\prime}) .
\end{equation}
We choose our operator basis and renormalization scheme to agree with
\cite{Buras:2000if} upon the substitution $d \to b$, $\bar s
\to \bar c$, $\bar u \to \bar s$:
\begin{eqnarray}  \label{eq:basiscc}
Q_1^c &=& (\bar c_L^i \gamma_\mu b_L^j) (\bar s_L^j \gamma^\mu c_L^i) , 
\qquad 
Q_2^c = (\bar c_L^i \gamma_\mu b_L^i) (\bar s_L^j \gamma^\mu c_L^j) ,
\nonumber \\[2mm]
Q_3^c &=& (\bar c_R^i b_L^j) (\bar s_L^j c_R^i) ,
\qquad \qquad 
Q_4^c = (\bar c_R^i  b_L^i) (\bar s_L^j  c_R^j) ,
\nonumber \\[2mm]
Q_5^c &=& (\bar c_R^i \gamma_\mu b_R^j) (\bar s_L^j \gamma^\mu c_L^i) , 
\qquad 
Q_6^c = (\bar c_R^i \gamma_\mu b_R^i) (\bar s_L^j \gamma^\mu c_L^j) ,
\nonumber \\[2mm]
Q_7^c &=& (\bar c_L^i b_R^j) (\bar s_L^j c_R^i) ,
\qquad \qquad 
Q_8^c = (\bar c_L^i  b_R^i) (\bar s_L^j  c_R^j) ,
\nonumber \\[2mm]
Q_9^c &=& (\bar c_L^i \sigma_{\mu\nu} b_R^j) (\bar s_L^j \sigma^{\mu\nu} c_R^i) ,
\quad 
Q_{10}^c = (\bar c_L^i  \sigma_{\mu\nu}  b_R^i) (\bar s_L^j \sigma^{\mu\nu} c_R^j) .
\nonumber \\
\label{opbasis}
\end{eqnarray}
The  $Q_i^{c\prime}$  are obtained
by changing all the quark chiralities. We leave a discussion of
such ``right-handed current'' effects
for future work \cite{JKLL} and
    discard the $Q_i^{c\prime}$ below.
    We split the Wilson coefficients into
SM and BSM parts,
\begin{equation}  \label{eq:wcsplit}
C_i^c(\mu) = C_i^{c, \rm SM}(\mu) + \Delta C_i(\mu) ,
\end{equation}
where $C_i^{c, \rm SM} = 0$ except for $i=1,2$ and $\mu$ is the
renormalization scale.

\section{Rare $B$ decays}
\begin{figure}
\vspace{0pt}
    \includegraphics[width=32mm]{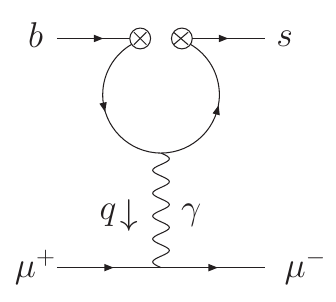}
    \hspace{0.5cm} 
\includegraphics[width=45mm]{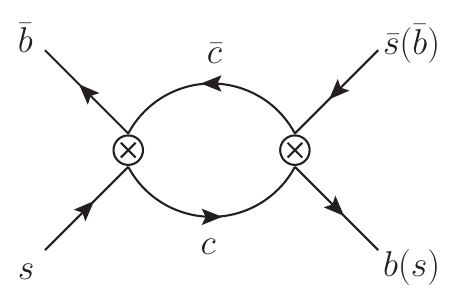}
\caption{
Leading CBSM contributions to rare decays (left), and to width
difference $\Delta \Gamma_s$ and lifetime ratio $\tau (B_s) / \tau (B_d)$
(right). 
\label{fig:diagrams}
}
\end{figure}

The leading-order (LO), one-loop CBSM effects in radiative and
rare semileptonic decays
may be expressed through ``effective'' Wilson coefficient
contributions $\Delta C_9^{\rm eff}(q^2)$ and $\Delta
C_7^{\rm eff}(q^2)$ in an  effective local Hamiltonian 
\begin{eqnarray}\label{eq:heffccdecay}
    \mathcal{H}_{\rm eff}^{rsl} &=& - \frac{4G_{F}}{\sqrt{2}} V_{ts}^* V_{tb}
        \left(C_7^{\rm eff}(q^2) Q_{7\gamma} + C_9^{\rm eff}(q^2) Q_{9V}
    \right) ,
\end{eqnarray}
where $q^2$ is the dilepton mass and
$$
Q_{7\gamma} =  \frac{e\, m_b}{16\pi^{2}}(\bar s_L \sigma_{\mu\nu} b_R) F^{\mu\nu},
\;\;\;
Q_{9V} =  \frac{\alpha}{4\pi}(\bar s_L \gamma_\mu b_L) (\bar \ell \gamma^\mu \ell).
$$
For $q^2$ small (in particular, well below the charm resonances),
$\Delta C_9^{\rm eff}(q^2)$ and $\Delta
C_7^{\rm eff}(q^2)$ govern the theoretical predictions
for both exclusive ($B \to K^{(*)} \ell^+ \ell^-,
B_s \to \phi \ell^+ \ell^-$, etc.) and inclusive $B  \to X_s \ell^+ \ell^-$ decay,
up  to ${\cal O}(\alpha_s)$ QCD corrections and power corrections to
the heavy-quark limit that we neglect in
our leading-order analysis. Similarly, $\Delta C_7^{\rm eff}(0)$ determines
radiative $B$-decay rates.
We will neglect the small CKM combination $V_{us}^* V_{ub}$, implying
$V_{cs}^* V_{cb}= - V_{ts}^* V_{tb}$, and focus on real (CP-conserving)
values for the $C_i^c$.
From the diagram shown in Fig.~\ref{fig:diagrams} (left) we then  obtain
\begin{eqnarray}
\label{eq:C9}
\Delta C_9^{\mathrm{eff}}(q^2)
&=& \left( C_{1,2}^c - \frac{C_{3,4}^c}{2} \right) h - \frac29 C_{3,4}^c \, ,
\\
\label{eq:C7}
\Delta C_7^{\mathrm{eff}}(q^2)
&=&\frac{m_{c}}{m_{b}} \! \left[ \!
\left( 4 C_{9,10}^c -\! C_{7,8}^c \right) y +\! \frac{4 C_{5,6}^c - C_{7,8}^c}{6} \right],
\end{eqnarray}
with $C_{x,y}^c = 3 \Delta C_x + \Delta C_y$ and the loop functions 
\begin{align}
    h(q^2, m_c, \mu) &= -\frac{4}{9} \left[ \ln \frac{m_c^2}{\mu^2} -
        \frac{2}{3}  + (2 + z) a(z) - z  \right],
\\
    y(q^2, m_c, \mu) &=  -\frac{1}{3} \left[ \ln \frac{m_c^2}{\mu^2} 
    - \frac32 + 2 a(z) \right],
\end{align}
where $a(z)= \sqrt{|z-1|} \arctan \frac{1}{\sqrt{z -1}}$ and
    $z = 4 m_c^2/q^2$.
Our numerical evaluation employs the charm pole mass.

We note that only the four Wilson coefficients $\Delta C_{1 \dots 4}$ enter
$\Delta C_9^{\rm eff}(q^2)$. Conversely, $\Delta C_7^{\rm eff}(q^2)$
is given in terms of the other six Wilson coefficients $\Delta C_{5 \dots 10}$.
The appearance of a one-loop, $q^2$-dependent contribution
to $C_7^{\rm eff}$ is a novel feature in the CBSM scenario.
Numerically, the loop function $a(z)$ equals one at $q^2 = 0$
and vanishes at $q^2 = (2 m_c)^2 $.
The constant terms and the logarithm accompanying $y(q^2, m_c)$
partially cancel the contribution from $a(z)$ and they introduce a sizable dependence on the renormalization scale $\mu$ and 
the charm quark mass.
Since a shift of  $\Delta C_7^{\rm eff}(q^2)$
is strongly constrained by the measured $B \to X_s \gamma$
decay rate, we do not consider the coefficients
$\Delta C_{5 \dots 10}$ in the remainder and focus on the four coefficients
$\Delta C_{1 \dots 4}$, which do not contribute to $B \to X_s \gamma$
at 1-loop order. Higher-order contributions can be important if 
new physics generates $\Delta C_i$ at the weak scale
or beyond, as is typically expected.
In this case large
logarithms $\ln M/m_B$ occur, requiring resummation.
To leading-logarithmic accuracy, we find
\begin{eqnarray}
    \Delta C_7^{\rm eff} &=& 
    0.02  \Delta C_1 \!-\!0.19 \Delta C_2 
    \!-\!0.01 \Delta C_3 \!-\!0.13  \Delta C_4 ,
\label{eq:c7hs}
    \\
    \Delta C_9^{\rm eff} &=& 
8.48  \Delta C_1 \!+\! 1.96  \Delta C_2
\!-\!4.24 \Delta C_3 \!-\!1.91 \Delta C_4 , \qquad
\label{eq:c9hs}
\end{eqnarray}
if $\Delta C_i$ are understood to be renormalized at $\mu=M_W$
and $\Delta C_{7,9}^{\rm eff}$ at $\mu=4.2$ GeV.
It is clear that $\Delta C_1$
and $\Delta C_3$ contribute (strongly) to rare semileptonic decay
but only weakly to $B \to X_s \gamma$.

\section{Mixing and lifetime observables}
A distinctive feature of the CBSM scenario is that nonzero 
$\Delta C_i$ affect not only radiative and rare semileptonic decays, but also
tree-level hadronic $b \to c \bar c s$ transitions. While the
theoretical control over exclusive $b \to c \bar c s$ modes
is very limited at present,  the decay width difference $\Delta
\Gamma_s$ and the lifetime ratio $\tau (B_s) / \tau (B_d)$ stand out
as being calculable in a heavy-quark expansion
\cite{Khoze:1983yp,*Shifman:1984wx,*Bigi:1992su}; see Fig.\
\ref{fig:diagrams} (right).
For both observables, the heavy-quark expansion
gives rise to an operator product expansion
in terms of local $\Delta B=2$ (for the width difference) or
$\Delta B=0$ (for the lifetime ratio) operators. The formalism is reviewed
in \cite{Lenz:2014jha} and applies to both SM and
CBSM contributions.
For the $B_s$ width difference, we have \cite{Artuso:2015swg}
$
\Delta \Gamma_s =
        2 |\Gamma_{12}^{s,\rm SM} + \Gamma_{12}^{c \bar c}| \cos \phi_{12}^s ,
$
where the phase $ \phi_{12}^s$ is small.
Neglecting the strange-quark mass, we find
\begin{equation}
\begin{split}
    \Gamma_{12}^{cc} &= - G_F^2 (V_{cs}^* V_{cb}^{})^2m_b^2  M_{B_s} f_{B_s}^2 \frac{\sqrt{1\!-\!4 x_c^2}}{576\pi} 
    \times \\
    &\Bigg\{ \Big[
    16 (1 - x_c^2) ( 4 C_2^{c,2} + C_4^{c,2})+ 8 (1-4x_c^2)  \times
    \\
    & 
    (12 C_1^{c,2} + 8 C_1^c C_2^c + 2 C_3^c C_4^c + 3 C_3^{c,2} ) - 192 x_c^2 \times
    \\
    & 
    (3 C_1^c C_3^c + C_1^c C_4^c + C_2^c C_3^c + C_2^c C_4^c) \Big] B+ 2 (  1 + 2 x_c^2) \times
    \\
    & 
    (4 C_2^{c,2} \! - 8  C_1^c C_2^c -12  C_1^{c,2}  \!- 3  C_3^{c,2}  \!- 2 C_3^c C_4^c + C_4^{c,2})\tilde{B_S^\prime} 
    \Bigg\} ,
\end{split}
\label{eq:Gamma12}
\end{equation}
with $x_c = m_c/m_b$. $B$,  $\tilde B_S'$ are defined through 
\begin{align}
    \langle B_s | (\bar s_L \gamma_\mu b_L) (\bar s_L \gamma^\mu b_L) \bar B_s \rangle
    &= \frac{2}{3} M^2_{B_S} f_{B_s}^2 B,
\\
    \langle  B_s | (\bar s_L^i b_R^j) (\bar s_L^j b^i_R) |\bar B_s \rangle
    &=\frac{1}{12} M_{B_s}^2 f_{B_s}^2 \tilde B_S' ,
\end{align}
with values taken from \cite{Aoki:2013ldr}.
For our numerical evaluation of $ \Gamma_{12}^{cc}$, we  split the
    Wilson coefficients according to (\ref{eq:wcsplit}),
subtract from the LO expression (\ref{eq:Gamma12}) the pure SM contribution 
and add the NLO SM expressions from \cite{Beneke:1996gn,*Dighe:2001gc,*Beneke:1998sy,*Beneke:2003az,*Ciuchini:2003ww,*Lenz:2006hd}.
In general, a modification of $\Gamma_{12}^{cc}$ also affects
the semi-leptonic CP asymmetries. However, since we consider
CP-conserving new physics in this paper 
and since the corresponding experimental uncertainties are
still large, the semi-leptonic asymmetries will not
lead to an additional constraint.

In a similar manner, for the the lifetime ratio, we find
\begin{align}
    \frac{\tau_{B_s}}{\tau_{B_d}} = \left( \frac{\tau_{B_s}}{\tau_{B_d}}
    \right)_{\rm SM} + \left( \frac{\tau_{B_s}}{\tau_{B_d}} \right)_{\rm NP} ,
\end{align}
where the SM contribution is
taken from \cite{Franco:2002fc}
and
\begin{eqnarray}
\lefteqn{\left( \frac{\tau_{B_s}}{\tau_{B_d}} \right)_{\!\!\rm NP}\!\!\!\!
    = G_F^2 |V_{cb} V_{cs}|^2 m_b^2 M_{B_s} f_{B_s}^2 \tau_{B_s}
    \frac{\sqrt{1\!-\!4 x_c^2}}{144\pi}  \times } \qquad \quad & & 
    \nonumber \\
    & &\Bigg\{
    (1- x_c^2) \Big[(4 C_{1,2}^{c,2} + C_{3,4}^{c,2})B_1 +6 (4 C_{2}^{c,2} + C_{4}^{c,2}) \epsilon_1 \Big]
    \nonumber \\
& &
    -12 x_c^2 \Big( C_{1,2}^c C_{3,4}^c  B_1  + 6 C_{2}^c C_{4}^c  \epsilon_1 \Big)   -  \frac{M_{B_s}^2( 1 + 2 x_c^2)}{(m_b + m_s)^2} \times
    \nonumber \\
    &  &  \Big[ 
    (4 C_{1,2}^{c,2} + C_{3,4}^{c,2} ) B_2 + 6 (4 C_{2}^{c,2} + C_{4}^{c,2} ) \epsilon_2
    \Big]
\Bigg\}, 
\end{eqnarray}
subtracting the SM part and defining
$B_1$, $B_2$, $\epsilon_1$, $\epsilon_2$ as
\begin{align}
    & \langle B_s | (\bar b_L \gamma_\mu s_L) (\bar s_L \gamma^\mu b_L) | B_s \rangle
    = \frac{1}{4} f_{B_s}^2 M_{B_s}^2 B_1,
\\
    & \langle B_s | (\bar b_R s_L) (\bar s_L b_R) | B_s \rangle
    = \frac{1}{4} \left[ \frac{M_{B_s}}{(m_b + m_s)} \right]^2 f_{B_s}^2 M_{B_s}^2 B_2,
    \\
    &\langle B_s | (\bar b_L \gamma_\mu T^A s_L) (\bar s_L \gamma^\mu T^A b_L) | B_s \rangle
    = \frac{1}{4} f_{B_s}^2 M_{B_s}^2 \epsilon_1,
    \\
&  \langle B_s | (\bar b_R T^A s_L) (\bar s_L T^A b_R) | B_s \rangle
    = \frac{1}{4} \left[ \frac{M_{B_s}}{(m_b + m_s)} \right]^2 f_{B_s}^2 M_{B_s}^2 \epsilon_2 ,
\end{align}
with values taken from \cite{Kirk:2017juj}.
We interpret the quark masses as $\overline{\rm MS}$
parameters at $\mu=4.2$ GeV.

\section{Rare decays versus lifetimes---low-scale scenario}
\begin{figure}
\begin{tabular}{cc}
    \includegraphics[width=43mm]{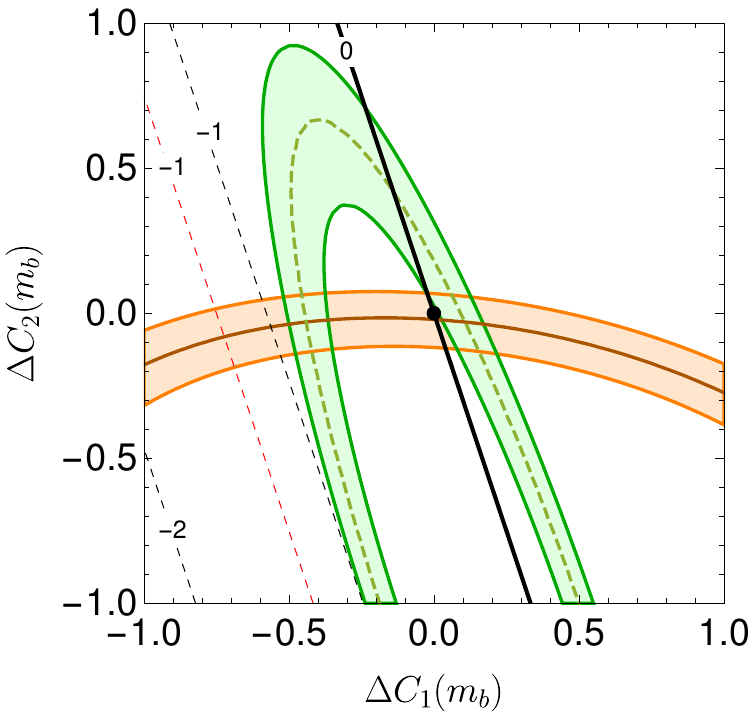}\hspace{0.2cm} &
    \includegraphics[width=39mm]{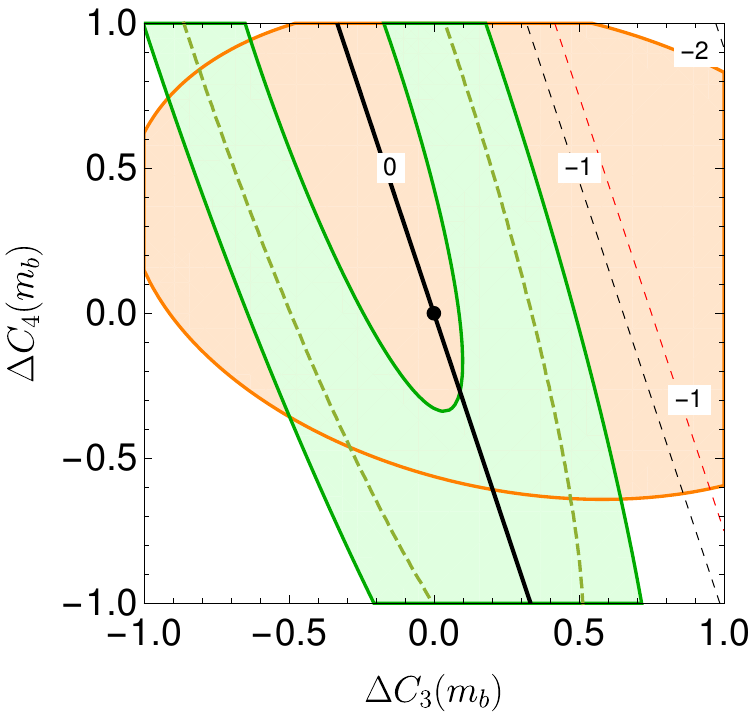}
\end{tabular}
\caption{
    Mixing observables versus rare decays in the CBSM scenario.
    Left: $(\Delta C_1, \Delta C_2)$ plane,
    Right: $(\Delta C_3, \Delta  C_4)$ plane. In each case, all Wilson
        coefficients are renormalized at $\mu = 4.2$ GeV and those
        not corresponding to either axis set to zero.
        The black dot corresponds to the SM, i.e. $\Delta C_i = 0$.
        The measured central value for the width difference is shown as brown
(solid) line together with the $1\sigma$ allowed region. The
lifetime ratio measurement is depicted as green (dashed) line and band.
Overlaid are contours of $\Delta C_9^{\rm eff}(5 {\rm GeV}^2) =
-1,-2$ (black, dashed) and $\Delta C_9^{\rm eff}(2 {\rm GeV}^2) =
-1,-2$ (red, dotted), as computed from (\ref{eq:C9}), and
of $\Delta C_9^{\rm eff} = 0$ (black, solid).
\label{fig:obslow}
}
\end{figure}

We are now in a position to confront the CBSM scenario with rare decay and
mixing observables, as long as we consider renormalization scales $\mu \sim m_B$.
Then the logarithms inside the $h$ function entering (\ref{eq:C9}) are small
and our leading-order calculation should be accurate. Such a scenario is
directly applicable if the mass scale $M$ of the physics
generating the $\Delta C_i$ is not too far above $m_B$, such that
$\ln (M/m_B)$ is small.
Fig.\ \ref{fig:obslow} (left) shows the
experimental $1\sigma$ allowed regions for the width difference
and lifetime ratio
(from the web update of \cite{Amhis:2016xyh})
in the $(\Delta C_1, \Delta C_2)$ plane.
The central values are attained on the brown (solid) and green
(dashed) curves, respectively.
The measured lifetime ratio and the width difference measurement
    can be simultaneously accommodated for different values  of the Wilson coefficients:
    in the $\Delta C_1$-$\Delta C_2$ plane, we find the SM solution, as well as a solution
    around $\Delta C_1= -0.5$ and $\Delta C_2 \approx 0$.
    In the $\Delta C_3$-$\Delta C_4$ plane, we have a relatively broad allowed range, roughly covering
    the interval $[-0.9,+0.7]$ for $\Delta C_3$ and $[-0.6,+1.1]$ for $\Delta C_4$. For further conclusions,
    a considerably higher precision in experiment and theory is required for $\Delta \Gamma_s$
    and $\tau_{B_s} / \tau_{B_d}$.
Also shown in the plot are contour lines for the contribution
to the effective semileptonic coefficient $\Delta C_9^{\rm eff}(q^2)$,
both for  $q^2 = 2\, {\rm GeV}^2$ and $q^2 = 5\, {\rm GeV}^2$ 
We see that sizable negative shifts are possible while respecting
the measured width difference
and the lifetime ratio.
For example,
a shift $\Delta C_9^{\rm eff} \sim -1$ as data may suggest could be
achieved through $\Delta C_1 \sim -0.5$ alone.
Such a value for $\Delta C_1$ may well be
consistent with CP-conserving exclusive $b \to c \bar c s$ decay data,
where no accurate theoretical predictions exist.
On the other hand, $\Delta C_9^{\rm eff}$ only exhibits a mild $q^2$-dependence.
Distinguishing this
from possible long-distance contributions would require substantial
progress on the theoretical understanding of the latter.

We can also consider other Wilson coefficients, such as the pair
$(\Delta C_3, \Delta C_4)$ (right panel in Fig.\ \ref{fig:obslow}). A shift
$\Delta C_9^{\rm eff} \sim -1$ is equally possible and consistent with the width
difference, requiring only $\Delta C_3 \sim 0.5$.

\section{High-scale scenario and RGE}
\subsection{RG enhancement of $\Delta C_9^{\rm eff}$}

If the CBSM operators are generated at a high scale then large logarithms
$\ln M/m_B$ appear. Their resummation is achieved by evolving the
initial (matching) conditions $C_i(\mu_0 \sim M)$
to a scale $\mu \sim M_B$
according to the coupled renormalization-group equations (RGE),
\begin{equation}
    \mu \frac{d C_j}{d\mu}(\mu) = \gamma_{ij}(\mu) C_i(\mu) ,
\end{equation}
where $\gamma_{ij}$ is the anomalous-dimension matrix.
    As is well known, the operators $Q_i^c$ mix not
    only with $Q_7$ and $Q_9$,  but also with the 4 QCD penguin
    operators $P_{3\dots 6}$ and the chromodipole operator $Q_{8g}$
    (defined as in \cite{Chetyrkin:1996vx}), which in turn mix into $Q_7$.
    Hence the index $j$ runs over
    11 operators with $\Delta B=-\Delta S=1$ flavor quantum numbers in order
    to account for all contributions to $C_7(\mu)$ that are proportional
    to $\Delta C_i(\mu_0)$. 
    Most entries of $\gamma_{ij}$
are known at LO \cite{Gaillard:1974nj,*Altarelli:1974exa,*Gilman:1979bc,*Shifman:1976ge,Gilman:1979ud,Guberina:1979ix,
Chetyrkin:1996vx,
Ciuchini:1993ks,*Ciuchini:1994xa,*Ciuchini:1993fk,Buras:2000if,
Bertolini:1986th,*Grinstein:1987vj,*Grinstein:1990tj,*Misiak:1991dj,
Gilman:1979ud,Chetyrkin:1996vx,Chetyrkin:1997fm};
our novel results are ($i=3,4$)
$$ \gamma^{(0)}_{Q_i^c \tilde Q_9} \! = \!
    \left(\frac{4}{3}, \frac{4}{9} \right)_{\!\!i} \!\!,
\;\;
    \gamma^{(0)}_{Q_i^c P_4} \! = \! \left( 0 , -\frac{2}{3} \right)_{\!\!i} \! ,
    \;\;
    \gamma_{Q_i^c Q_7}^{\rm eff(0)} \! = \! \left(0, \frac{224}{81} \right)_{\!\!i} \!\! ,
    $$
where $\tilde Q_9=(4\pi/\alpha_s) Q_{9V}(\mu)$
and $\gamma_{Q_i^c Q_7}^{\rm eff(0)}$ requires a two-loop
calculation. (See appendix for further technical information.)
Solving the RGE for $\mu_0 = M_W$, $\mu = 4.2$ GeV,
and $\alpha_s(M_Z) = 0.1181$, results in the CBSM contributions to
$\Delta C_7^{\rm eff}$ and $\Delta C_9^{\rm eff}$ in (\ref{eq:c7hs}),(\ref{eq:c9hs})
as well as
\begin{equation}   \label{eq:rgetrunc}
    \left(  \!\! \begin{array}{c}
            \Delta C_1(\mu) \\ \Delta C_2(\mu) \\ \Delta C_3(\mu) \\ \Delta C_4(\mu) 
        \end{array} \!\!\right) \!\!=\!\!
\left(\!\!\! \begin{array}{cccc}
    1.12  & -0.27 & 0     & 0  \\
    -0.27 & 1.12  & 0     & 0 \\
    0     & 0     & 0.92  & 0  \\
    0     & 0     & 0.33  & 1.91
\end{array}\! \right) \!\!\!
\left(\!\! \begin{array}{c} 
            \Delta C_1(\mu_0) \\ \Delta C_2(\mu_0) \\ \Delta C_3(\mu_0)
            \\ \Delta C_4(\mu_0)
        \end{array} \! \right) \!\!.
\end{equation}
A striking feature are the large coefficients in the
$\Delta C_9^{\rm eff}$ case, which are ${\cal O}(1/\alpha_s)$ in the logarithmic
counting. The largest coefficients appear for $\Delta C_1$ and $\Delta
C_3$, which at the same time practically do not mix into $C_7^{\rm eff}$. This means that
small values $\Delta C_1 \sim -0.1$ or $\Delta C_3\sim 0.2$  can generate
$\Delta C_9^{\rm eff}(\mu) \sim -1$ while having essentially no impact
on the $B \to X_s \gamma$ decay rate. Conversely, values for $\Delta
C_2$ or $\Delta C_4$
that lead to $\Delta C_9^{\rm eff} \sim -1$ lead to large effects in $C_7^{\rm eff}$
and $B \to X_s \gamma$.

\subsection{Phenomenology for high NP scale}
\begin{figure}
\begin{tabular}{cc}
    \includegraphics[width=41mm]{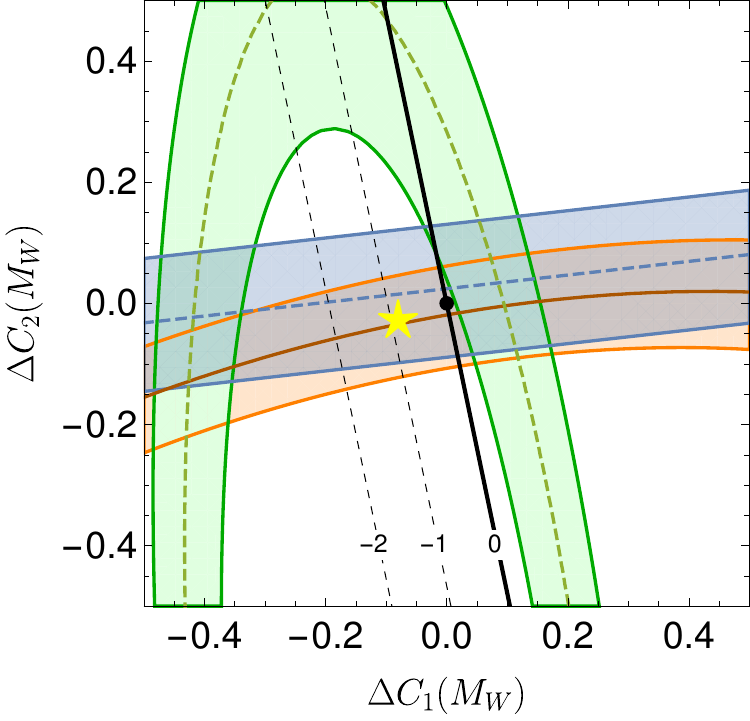}\hspace{0.2cm} &
    \includegraphics[width=41mm]{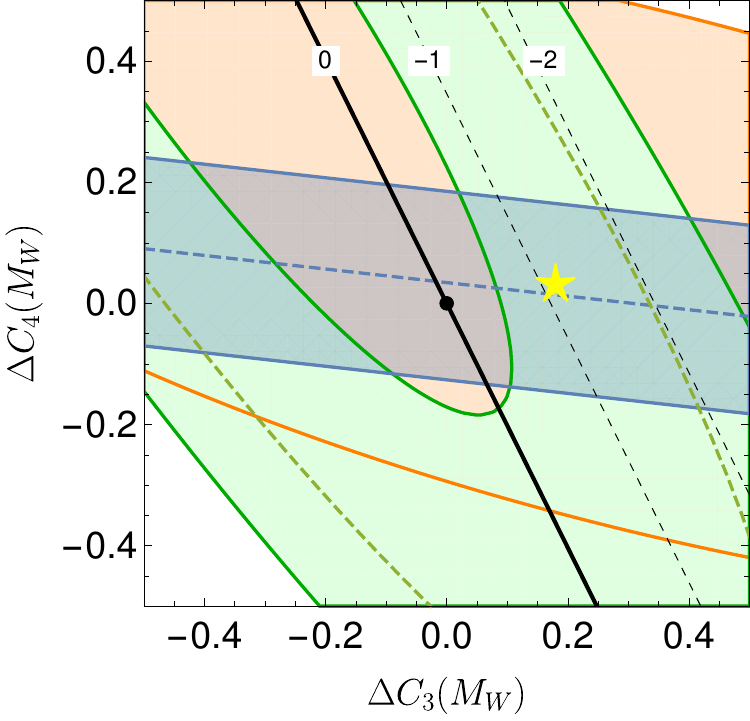}
    \\
    \includegraphics[width=41mm]{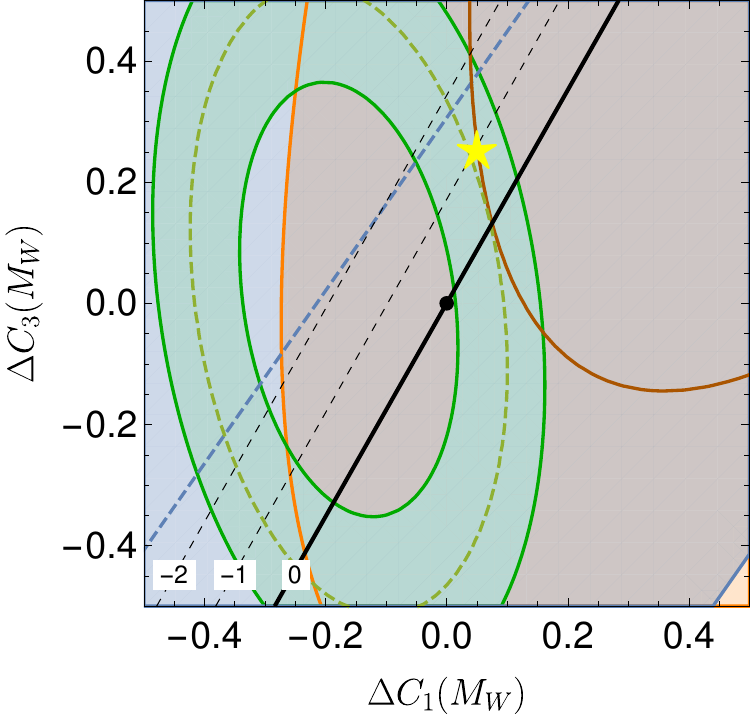}\hspace{0.2cm} &
    \includegraphics[width=41mm]{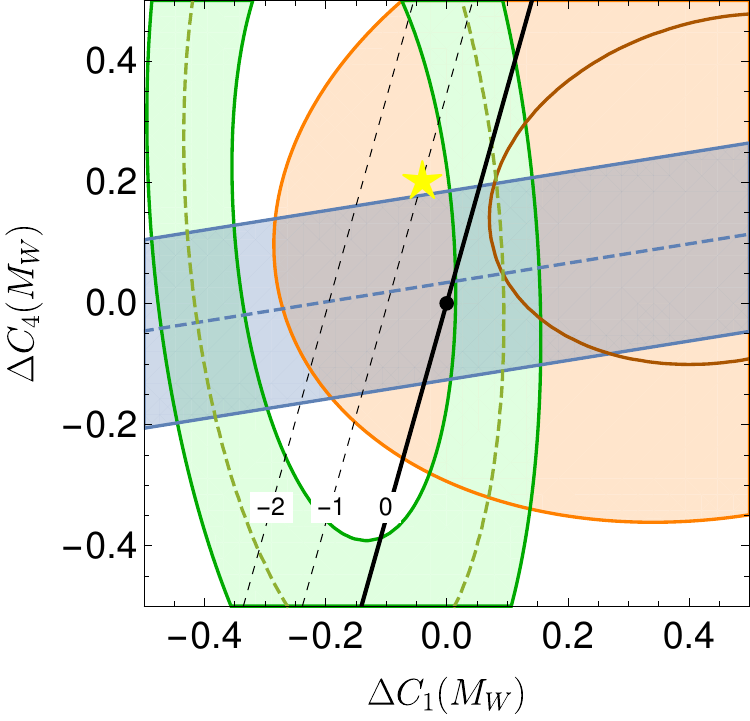}
    \\
    \includegraphics[width=41mm]{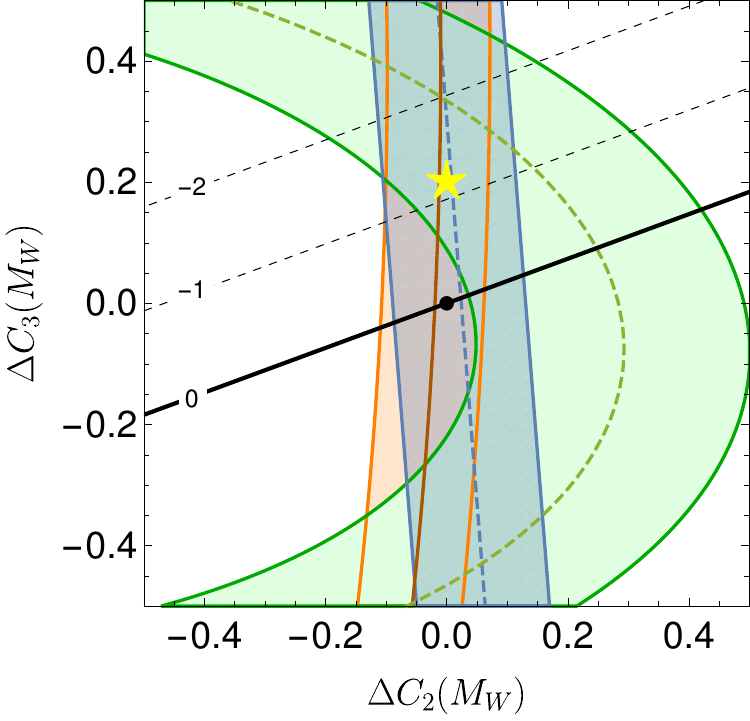}\hspace{0.2cm} &
    \includegraphics[width=41mm]{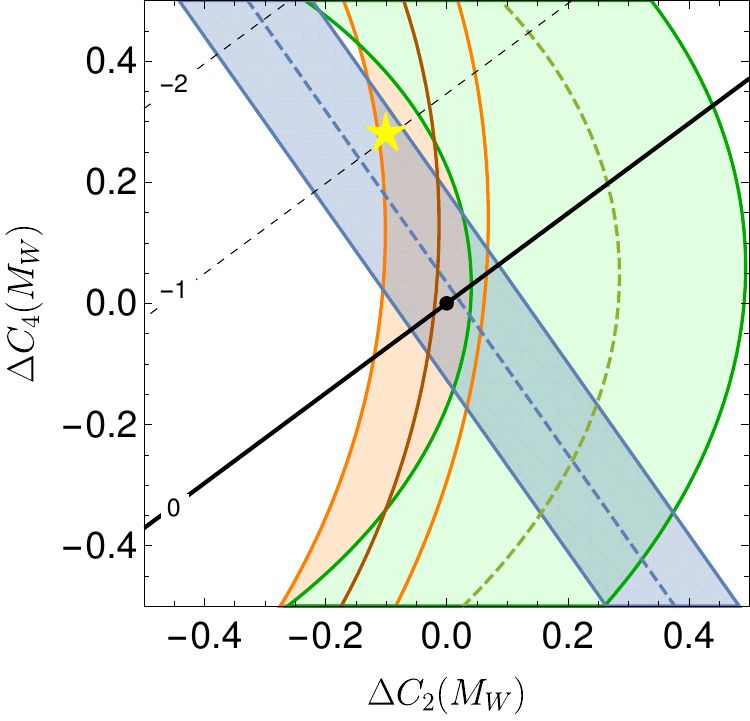}
\end{tabular}
\caption{
    Mixing observables versus rare decays, for $\Delta C_i$
    renormalized at $\mu_0=M_W$. Color coding as in Fig.~\ref{fig:obslow},
    $B \to X_s \gamma$ constraint shown in addition (straight blue bands).
\label{fig:obshigh}
}
\end{figure}
The situation in various two-parameter planes is depicted 
in Fig.~\ref{fig:obshigh}, where the $1\sigma$ constraint
from $B \to X_s \gamma$ is shown as blue, straight bands.
(We implement it by splitting $BR(B \to X_s \gamma)$ into SM and BSM
parts and employ the numerical result and theory error from \cite{Misiak:2015xwa} for
the former. The experimental result is taken from the web update of
\cite{Amhis:2016xyh}.)
The top row corresponds to Fig.\ \ref{fig:obslow}, but contours of
given $\Delta C_9$ lie much closer to the origin.
    All six panels testify to the fact that the SM is consistent
    with all data when leaving aside the question of rare semileptonic $B$ decays---the
    largest pull stems from the fact that the experimental value
    for $\tau_{B_s}/\tau_{B_d}$ is
    just under 1.5 standard deviations below the SM expectation,
    such that the black (SM) point is less than $0.5\sigma$ outside the
    green area.
    Our main question is now: can we have  a new contribution  $\Delta C_9^{\rm eff} \sim -1$ to rare semileptonic decays, while being consistent
    with the bounds stemming from $b \to s \gamma$, $\Delta \Gamma_s$ and $\tau_{B_s}/\tau_{B_d}$? This is clearly possible
    (indicated by the yellow star in the plots)
    if we have a new contribution $\Delta C_3 \approx 0.2$, see the three plots of the $\Delta C_i - \Delta C_3$ planes
    in Fig. \ref{fig:obshigh} (right on the top row, left on the middle row and left on the lower row). In these cases, the
    $\Delta C_9^{\rm eff} \sim -1$ solution is even favored compared to the SM solution.
    A joint effect in $\Delta C_2 \approx -0.1$ and $\Delta C_4 \approx 0.3$ can also accommodate our  desired scenario, see
    the right plot on the lower row, while new BSM effects in the pairs $\Delta C_1, \Delta C_2$ and $\Delta C_1, \Delta C_4$
    alone are less favored.
One could also consider three or all four $\Delta C_i$ simultaneously.
    
\subsection{Implications for UV physics}
    Our model-independent results are well suited to study the
rare $B$-decay and lifetime phenomenology of ultraviolet
(UV) completions of the Standard Model. Any such completion may include
extra UV contributions to $C_7(M)$ and $C_9(M)$, correlations with
other flavor observables, collider phenomenology, etc.; the
details are highly model-dependent and beyond the scope
of our model-independent analysis. Here we restrict ourselves
to some basic sanity checks.

Taking the case of $\Delta C_1(M) \sim -0.1$ corresponds to
a naive ultraviolet scale
$$
\Lambda \sim \left( \frac{4 G_F}{\sqrt{2}} |V_{cs}^* V_{cb}| \times 0.1 \right)^{-1/2}
    \sim 3\,\mbox{TeV} .
$$
This effective scale could arise in a weakly-coupled scenario
from tree-level exchange of new scalar or vector mediators,
or at loop level in addition from fermions; or the effective operator
could arise from strongly-coupled new physics. For a tree-level exchange,
$\Lambda \sim M/g_*$ where $g_* = \sqrt{g_1 g_2}$ is the geometric mean
of the relevant couplings. For weak coupling $g_* \sim 1$, this then
gives $M \sim 3$ TeV. Particles of such mass are certainly allowed by collider
searches if they do not couple (or only sufficiently weakly)
to leptons and first-generation quarks.
Multi-TeV weakly coupled particles also generically are not in  violation
of electroweak precision tests of the SM.
Loop-level mediation would require mediators close to the weak scale
which may be problematic and would require a specific
investigation; this is of course unsurprising given
that $b \to c \bar c s$ transitions are mediated at tree level in the SM.
The same would be true in a BSM scenario that mimics the flavor suppressions
in the SM (such as MFV models). Conversely, in a strongly-coupled scenario
we would have $M \sim g_* \Lambda \sim 4 \pi \Lambda \sim 30$ TeV. This
is again safe from generic collider and precision constraints, and a
model-specific analysis would be required to say more.

Finally, as all CBSM effects are lepton-flavor-universal,
they cannot on their own account for departures  of the lepton
flavor universality parameters $R_{K^{(*)}}$ \cite{Hiller:2003js} from
the SM values as suggested
by current experimental measurements \cite{Aaij:2014ora,*Aaij:2017vbb,*Aaij:2017tyk}. However, even if those departures are real, they
may still be caused by direct
UV contributions to $\Delta C_9$. For example, as shown in \cite{Geng:2017svp},
a scenario with a muon-specific contribution
$\Delta C_9^{\mu} = - \Delta C_{10}^\mu \sim -0.6$ and in addition a
lepton-universal contribution $\Delta C_9 \sim -0.6$, which may have a
CBSM origin, is perfectly consistent with all rare-$B$-decay data,
and in fact marginally preferred.

\section{Prospects and summary}
\begin{figure}
\begin{tabular}{cc}
    \includegraphics[width=41mm]{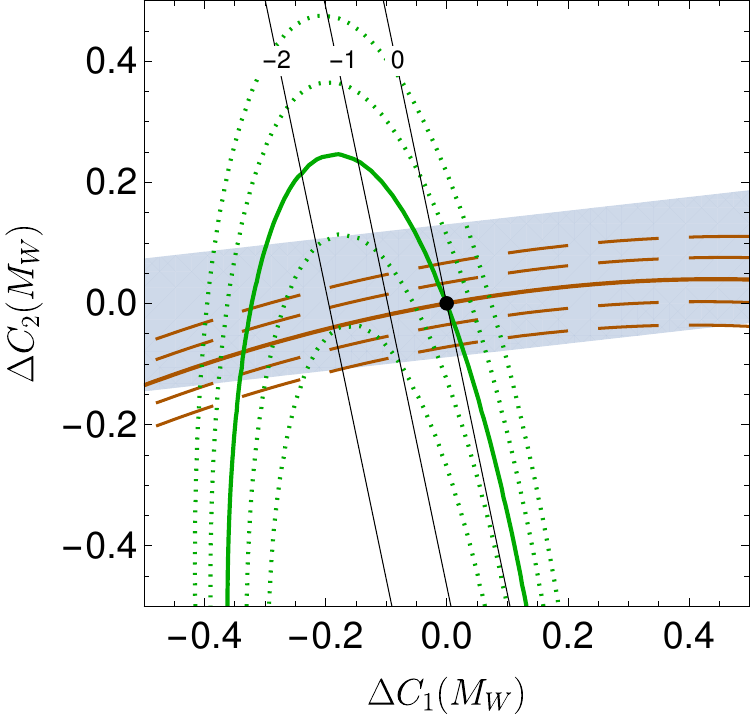}\hspace{0.2cm} &
    \includegraphics[width=41mm]{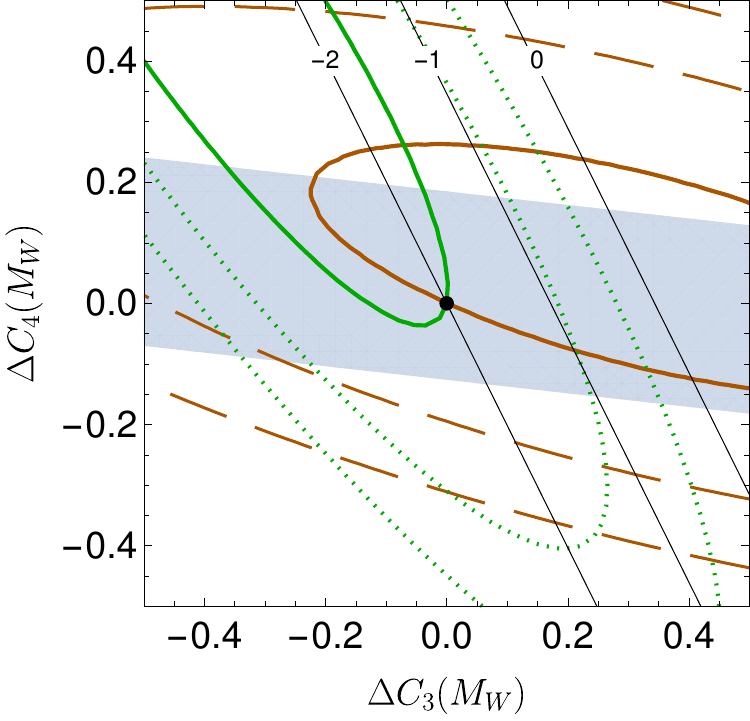}
\end{tabular}
\vspace*{-0.3cm}
\caption{
Future prospects for mixing observables. Dashed: contours of constant
width difference, dotted: contours of constant lifetime ratio. See text for discussion.
\label{fig:prospects}
}
\end{figure}
The preceding discussion suggests that
a precise knowledge of width difference and lifetime ratio,
as well as $BR(B \to X_s \gamma)$,
    can have the potential to identify and discriminate between different CBSM scenarios, or
rule them out altogether.
This is illustrated in Fig.\ \ref{fig:prospects}, showing contour values for
future precision both in mixing and lifetime observables.
In each panel, the solid (brown and green) contours correspond to
the SM central values of the width difference and lifetime ratio
(respectively).  The  spacing of the accompanying contours is such that the area
between any two neighboring contours
corresponds to a prospective $1\sigma$-region, assuming a
combined (theoretical and experimental) error on the lifetime ratio of
0.001 and a combined error on $\Delta \Gamma_s$ of $5\%$.
The assumed future errors are ambitious but seem
feasible with expected experimental and theoretical progress.
Overlaid is the (current) $B \to X_s \gamma$ constraint (blue).
    The figure indicates that a discrimination between the SM and the scenario where
    $\Delta C_9 \approx -1$, while $BR(B \to X_s \gamma)$ is SM-like is clearly possible.
    A crucial role is played by the lifetime ratio $\tau_{B_s}/\tau_{B_d}$: in e.g. the
    $\Delta C_3-\Delta C_4$ case a 1 $\sigma$ deviation of the lifetime ratio almost coincides with the
    $\Delta C_9  =-1$ contour line; a further precise determination of $\Delta \Gamma_s$ could
    then identify the point on this line chosen by nature.
Further progress on $B \to X_s \gamma$ in the Belle II era
would provide complementary information.

In summary, we have given a comprehensive, model-independent
analysis of BSM effects in partonic
    $b \to c \bar c s$ transitions (CBSM scenario)
in the CP conserving case,
focusing on those observables that can be computed in a heavy-quark
expansion. An effect in
rare semileptonic $B$ decays compatible with
hints from current LHCb and $B$-factory
data can be generated, while satisfying the $B \to X_s
\gamma$ constraint. It can originate from different combinations
of $b \to c \bar c s$ operators. The required Wilson coefficients are
so small that constraints from $B$ decays into charm are not
effective, particularly if new physics enters at a high scale; then large
renormalization-group enhancements are present.
Likewise, there are no obvious model-independent conflicts with
    collider searches or electroweak precision observables.
A  more precise measurement of mixing observables and lifetime ratios, at a level
achievable at LHCb, may be able to confirm
(or rule out) the CBSM scenario, and to discriminate between different BSM
couplings.
    Finally, all CBSM effects are lepton-flavor-universal; the current $R_K$
    and $R_{K^*}$ anomalies would either have to be mismeasurements or
    require additional lepton-flavor-specific UV contribution to $C_9$; such
    a combined scenario has been shown elsewhere to be consistent with all
    rare $B$-decay data and also presents the most generic way for UV physics
    to affect rare decays.
With the stated caveats,
our conclusions are rather model independent. It would be interesting to construct
concrete UV realizations of the CBSM scenario,
which almost certainly will affect other observables
    in a correlated, but model-dependent manner.

\section{Acknowledgments}
We would like to thank C.\ Bobeth, P.\ Gambino, M.\ Gorbahn, and especially M.\ Misiak for discussions. This work
was supported by an IPPP Associateship. 
S.J. and K.L. acknowledge support
by STFC Consolidated Grant No. ST/L000504/1,
an STFC studentship, and a Weizmann
Institute ``Weizmann-UK Making Connections'' grant. A.L. and M.K. are supported
by the STFC IPPP grant.

    \section{Appendix: Technical aspects of the anomalous-dimension calculation}
Here we provide additional technical information regarding our results on
anomalous dimensions entering in the RGE (20).

A set of Wilson coefficients that contains $C_7$, $C_9$, and
$C^c_{1 \dots 4}$ and is closed under renormalization
necessarily also contains four QCD-penguin coefficients $C_{P_i}$ multiplying
the operators $P_{3 \dots 6}$ (we define them as in
\cite{Chetyrkin:1996vx})
and the chromodipole coefficient $C_{8g}$,
resulting in an $11 \times 11$ anomalous-dimension matrix
$\gamma$. If the rescaled semileptonic operator
$\tilde Q_9(\mu) = (4\pi/\alpha_s(\mu)) Q_{9V}(\mu)$ is used
then to leading order $\gamma_{ij}(\mu) = \alpha_s(\mu)/(4\pi) \gamma^{(0)}_{ij}$,
with constant $\gamma^{(0)}_{ij}$.
As is well known, this matrix is scheme-dependent already at LO
\cite{Ciuchini:1993ks,*Ciuchini:1994xa,*Ciuchini:1993fk}.
A scheme-independent matrix $\gamma^{\rm eff(0)}$
can be achieved by replacing $C_7$ and $C_8$
by the scheme-independent combinations
\begin{eqnarray}
    C_7^{\rm eff} &=& C_7 + \sum_i y_i C_i ,
\\
C_8^{\rm eff} &=& C_8 + \sum_i z_i C_i ,
\end{eqnarray}
where
\begin{eqnarray}
    \langle s \gamma | Q_i | b \rangle &=&
        y_i \langle s \gamma | Q_{7\gamma} | b \rangle, \\
    \langle s g | Q_i | b \rangle &=&
        z_i \langle s \gamma | Q_{8g} | b \rangle,
\end{eqnarray}
to lowest order and the sums run over all four-quark
operators. We find that $y_i$ and $z_i$ vanish for $Q_{1 \dots 4}^c$, leaving
only the known coefficients $y_{P_i} = (-1/3, -4/9, -20/3, -80/9)_i$
and $z_{P_i} = (1, -1/6, 20, -10/3)_i$  ($i=3 \dots
6$) \cite{Chetyrkin:1996vx}.
The BSM correction $\Delta C_9^{\rm eff}$ in (5),(10) coincides
with the (BSM correction to the) coefficient  $C_9$ of $Q_{9V}$ to LL accuracy.

Many of the
elements of $\gamma^{\rm eff(0)}$ are known
\cite{Gaillard:1974nj,*Altarelli:1974exa,*Gilman:1979bc,*Shifman:1976ge,Gilman:1979ud,Guberina:1979ix,Ciuchini:1993ks,*Ciuchini:1994xa,Buras:2000if},
except for
$\gamma^{\rm eff(0)}_{Q_{i}^c  Q_{7\gamma}}$,
$\gamma^{\rm eff(0)}_{Q_{i}^c Q_{8g}}$,
$\gamma^{\rm eff(0)}_{Q_i^c P_j}$,
and $\gamma^{\rm eff(0)}_{Q_{i}^c \tilde Q_9}$, for $i=3, 4$. 
The latter can be read off from the logarithmic terms in
(5), and the  mixing into $P_i$ follows
from substituting gauge
coupling and color factors in diagram Fig.~1
(left). This gives
$$ \gamma^{(0)}_{Q_i^c \tilde Q_9} =
    \left(-\frac{8}{3}, -\frac{8}{9}, \frac{4}{3}, \frac{4}{9} \right)_{\!\!i},
\quad
    \gamma^{(0)}_{Q_i^c P_4} = \left(0 , \frac{4}{3} , 0 , -\frac{2}{3} \right)_{\!\!i} ,
$$
for $i=1,2,3,4$, with the mixing into $C_{P_{3,5,6}}$ vanishing.
    
The leading mixing into $C_7^{\rm eff}$ arises at two loops
\cite{Bertolini:1986th,*Grinstein:1987vj,*Grinstein:1990tj,*Misiak:1991dj}
and is the technically
most challenging aspect of this work. Our calculation
employs the 1PI (off-shell) formalism and the method
of \cite{Chetyrkin:1997fm} for computing UV divergences, which involves
an infrared-regulator mass and the appearance of a set of
gauge-non-invariant counterterms.
The result is
$$
\gamma_{Q_i^c Q_7}^{\rm eff(0)} =
    \left( 0, \frac{416}{81}, 0, \frac{224}{81} \right)_i 
\qquad (i=1, 2,3,4) .
$$
Our stated results for $i=1,2$ agree with the results
in \cite{Gilman:1979ud,Chetyrkin:1996vx},
which constitutes a cross-check of our calculation.

We have not obtained the 2-loop mixing of $C_{3,4}^c$ into
$C_{8g}$ and set these anomalous dimension elements to zero.
For the case of $C_{1,2}^c$ where this mixing is known,
the impact of neglecting $\gamma^{\rm eff(0)}_{i8}$ on $\Delta C_7^{\rm eff}(\mu)$ is
small [the only change being $-0.19 \Delta C_2 \to -0.18\Delta C_2$ in (9)].
We expect a similarly small error in the case of $\Delta C_{3,4}$.

\end{document}